\documentclass{aastex631}

\usepackage{hyperref}

\accepted{October 2021}

\submitjournal{RNAAS}

\shorttitle{VeSPA}
\shortauthors{McMaster et al.}

\graphicspath{{./}{figures/}}

\begin{document}

\title{\texttt{VeSPA}: The SuperWASP Variable Star Photometry Archive}

\author[0000-0002-4785-7867]{Adam McMaster}
\affiliation{School of Physical Sciences, The Open University, Milton Keynes, MK7 6AA, UK}
\affiliation{DISCnet Centre for Doctoral Training, The Open University, Walton Hall, Milton Keynes, MK7 6AA, UK}

\author[0000-0001-7619-8269]{Andrew J. Norton}
\affiliation{School of Physical Sciences, The Open University, Milton Keynes, MK7 6AA, UK}

\author[0000-0003-0475-008X]{Hugh J. Dickinson}
\affiliation{School of Physical Sciences, The Open University, Milton Keynes, MK7 6AA, UK}

\author[0000-0003-4195-5068]{Heidi B. Thiemann}
\affiliation{School of Physical Sciences, The Open University, Milton Keynes, MK7 6AA, UK}
\affiliation{DISCnet Centre for Doctoral Training, The Open University, Walton Hall, Milton Keynes, MK7 6AA, UK}

\author[0000-0001-8670-8365]{Ulrich C. Kolb}
\affiliation{School of Physical Sciences, The Open University, Milton Keynes, MK7 6AA, UK}

\begin{abstract}

We present the first results from the \textit{SuperWASP Variable Stars} (SVS) citizen science project. The photometry archive of the Wide Angle Search for Planets has previously been searched for periodic variations and the results of this search formed the basis of the SVS project on the Zooniverse. The SVS project asks volunteers to visually inspect light curve plots and categorise each one according to a broad classification scheme. Results from the first two years of SVS have now been published online as the \textit{SuperWASP Variable Star Photometry Archive} (\texttt{VeSPA}). The archive can be browsed online, downloaded in full, or queried, filtered, and sorted to export a refined set of results. An interactive light curve viewer also allows any light curve to be folded at a user-defined period. Analysis of citizen science results and development of VeSPA features are both ongoing. Updated results will be published every six months.

\end{abstract}

\keywords{Variable stars (1761) --- Catalogs (205) --- Surveys (1671) --- Eclipsing binary stars (444) --- Pulsating variable stars (1307)}

\section{The SuperWASP Variable Star Catalogue}

The Wide Angle Search for Planets (\textit{WASP}) was a ground-based exoplanet search using the transit method to identify hot Jupiter exoplanet candidates from photometric light curves~\citep{pollacco_wasp_2006}. \textit{WASP} used two telescopes named \textit{SuperWASP North}, in La Palma, and \textit{SuperWASP South}, in South Africa. Between 2004 and 2013 the \textit{SuperWASP} cameras produced light curves of $\sim 31$ million sources, covering almost the entire sky (excluding the galactic plane) with magnitude brighter than $V \sim 15$~\citep{norton_zooniverse_2018}.

As noted in~\citet{norton_zooniverse_2018}, the \textit{SuperWASP} photometry archive is well suited to the study of variable stars given the long baseline and high cadence of observations. The 31 million \textit{SuperWASP} light curves were searched for periodic variations between $\sim 1$~hour and $\sim 1 $~year, using a power spectrum analysis using the \texttt{CLEAN} algorithm~\citep{roberts_time_1987} and phase dispersion minimisation~\citep{davies_improved_1990}; this resulted in 8 million periods which were detected by both methods (periods detected by only one method were not considered plausible)~\citep{thiemann_superwasp_2021}. Spurious periods corresponding to integer fractions or multiples of a sidereal day or lunar month were flagged and not included in the final catalogue~\citep{thiemann_superwasp_2021}. The light curves corresponding to the $\sim 1.6$~million resulting potential periods from $\sim 0.8$~million unique objects comprise the \textit{SuperWASP Periodicity Catalogue}. 

\textit{SuperWASP Variable Stars} (SVS) is a citizen science project that was set up on the Zooniverse~\citep{lintott_galaxy_2008} to enlist volunteers in classifying the resulting folded light curves~\citep{norton_zooniverse_2018}. The project launched in 2018 and in its first two years it received over a million classifications for 568,739 object-period combinations (i.e. unique folded light curve plots, where each source object can have multiple plots corresponding to multiple candidate periods), finding 2,560 previously unknown variable star candidates and categorising 190,068 candidates in total~\citep{thiemann_superwasp_2021}.

These results have now been published online in the SuperWASP Variable Star Photometry Archive (\texttt{VeSPA}).\footnote{Note that five candidates are not currently included due to a technical problem preventing the extraction of the raw photometry, so the total number of candidates in \texttt{VeSPA} is 190,063. We hope to include the missing candidates in a future data release.} See Table~\ref{tab:vespa} for a summary of the number of candidates published in each category.

\begin{deluxetable*}{cchlDlc}
\tablenum{1}
\tablecaption{The number of variable star candidates included in \texttt{VeSPA}. \label{tab:vespa}}
\tablewidth{0pt}
\tablehead{
\colhead{Candidate classification} & \colhead{Number of candidates}
}
\startdata
EA/EB & 29,882 \\
EW & 36,328 \\
Pulsator & 25,730 \\
Rotator & 56,582 \\
Unknown & 41,541 \\
Total & 190,063
\enddata
\end{deluxetable*}

\section{Accessing \texttt{VeSPA} and Using Data Exports}

The \texttt{VeSPA} website can be accessed via \href{https://www.superwasp.org/vespa/}{superwasp.org/vespa/}. The catalogue can be queried to perform a cone search using coordinates, catalogue names, and common names of objects. There are also options to filter and sort the results by magnitude, amplitude of variation, period length, and citizen science classification. An interactive light curve viewer allows users to fold each light curve at their own chosen period, with results displayed on the fly, in a manner similar to that used by the ASAS-SN Variable Stars Database~\citep{jayasinghe_asas-sn_2018}.

Data exports are available in CSV format. These can be generated via the \texttt{VeSPA} website by searching and filtering to get the desired set of results and then clicking ``Export as CSV". This will generate a \texttt{ZIP} archive containing the following three files:

\begin{itemize}
    \item \texttt{export.csv} -- The main data export, containing one row per folded light curve (i.e. multiple rows per source object).
    \item \texttt{params.yaml} -- A YAML-format copy of the search and filtering parameters which were used to generate the export, plus a data version number (which will be incremented with future data releases or changes to the export format).
    \item \texttt{fields.yaml} -- A YAML-format list of the columns included in the export with an English description of each one.
\end{itemize}

Python code examples for working with \texttt{VeSPA} data exports are available on GitHub at \href{https://github.com/adammcmaster/vespa-tools}{github.com/adammcmaster/vespa-tools}.

A copy of the full (unfiltered) data export for the first data release is available on Zenodo \citep{mcmaster_superwasp_2021}.

We encourage wide use of the \texttt{VeSPA} data and we welcome feedback as to how it might be improved or added to.

\section{Future Data Releases and Upcoming Features}

Our intention is to produce new data releases every six months. Each new data release will include new objects which have been classified by the Zooniverse volunteers, along with revised classifications for objects which have received additional classifications.

Each data release is versioned with a major and minor version number separated by a period. The major number will be incremented for new data releases, while the minor number will be used to indicate any minor technical modifications or corrections to the data exports. The current data release is 1.0. If we were to discover a formatting error, for example, we would correct this and release minor version 1.1. The next full data release would then be 2.0.

All data releases (full and minor) will be listed on the \href{https://www.superwasp.org/vespa/data-releases/}{data archive page on the \texttt{VeSPA} website (superwasp.org/vespa/data-releases/)}. The data archive also includes a full change log detailing what is included/changed in each release. The unfiltered export for each release can be downloaded from the data archive page, and these will also be uploaded to Zenodo. Only the latest release will be browsable/searchable via the \texttt{VeSPA} interface.

Development work on \texttt{VeSPA} and analysis of citizen science results are both ongoing. We are also exploring ways of speeding up citizen science classifications, and to this end we will be looking into augmenting human classifications with machine learning methods so that less human effort is required.

\section{Acknowledgements}

The SuperWASP project is currently funded and operated by Warwick University and Keele University, and was originally set up by Queen’s University Belfast, the Universities of Keele, St. Andrews and Leicester, the Open University, the Isaac Newton Group, the Instituto de Astrofisica de Canarias, the South African Astronomical Observatory and by STFC.

The Zooniverse project on SuperWASP Variable Stars is led by Andrew Norton (The Open University) and builds on work he has done with his former postgraduate students Les Thomas, Stan Payne, Marcus Lohr, Paul Greer, and Heidi Thiemann, and current postgraduate student Adam McMaster.

The Zooniverse project on SuperWASP Variable Stars was developed with the help of the ASTERICS Horizon2020 project. ASTERICS is supported by the European Commission Framework Programme Horizon 2020 Research and Innovation action under grant agreement n.653477

VeSPA was designed and developed by Adam McMaster as part of his postgraduate work. This work is funded by STFC, DISCnet, and the Open University Space SRA. Server infrastructure was funded by the Open University Space SRA.

\bibliography{references}{}
\bibliographystyle{aasjournal}

\end{document}